# Unveiling the mechanisms of the spin Hall effect in Ta

Edurne Sagasta[1], Yasutomo Omori[2], Saül Vélez[1,*], Roger Llopis[1], Christopher Tollan[1], Andrey Chuvilin[1,3], Luis E. Hueso[1,3], Martin Gradhand[4], YoshiChika Otani[2,5], Fèlix Casanova[1,5,†]

[1]CIC nanoGUNE, 20018 Donostia-San Sebastian, Basque Country, Spain
[2]Institute for Solid State Physics, University of Tokyo, Kashiwa, Chiba 277-8581, Japan
[3]IKERBASQUE, Basque Foundation for Science, 48013 Bilbao, Basque Country, Spain
[4]H. H. Wills Physics Laboratory, University of Bristol, Bristol BS8 1TL, United Kingdom
[5]RIKEN-CEMS, 2-1 Hirosawa, Wako, Saitama 351-0198, Japan
*present address: Department of Materials, ETH Zürich, 8093 Zürich, Switzerland
†Corresponding author: f.casanova@nanogune.eu

**Spin-to-charge current interconversions are widely exploited for the generation and detection of pure spin currents and are key ingredients for future spintronic devices including spin-orbit torques and spin-orbit logic circuits. In case of the spin Hall effect, different mechanisms contribute to the phenomenon and determining the leading contribution is peremptory for achieving the largest conversion efficiencies. Here, we experimentally demonstrate the dominance of the intrinsic mechanism of the spin Hall effect in highly-resistive Ta. We obtain an intrinsic spin Hall conductivity for β-Ta of -820±120 (ħ/e) Ω$^{-1}$cm$^{-1}$ from spin absorption experiments in a large set of lateral spin valve devices. The predominance of the intrinsic mechanism in Ta allows us to linearly enhance the spin Hall angle by tuning the resistivity of Ta, reaching up to -35±3 %, the largest reported value for a pure metal.**

Condensed matter systems with strong spin-orbit coupling (SOC) are extensively studied in the emerging field of spin-orbitronics due to the novel effects and functionalities originated from the interplay between the charge and the spin of electrons. The spin Hall effect (SHE) in heavy metals [1, 2] and Edelstein effect in Rashba interfaces [3, 4, 5] or in the Dirac surface states of topological insulators [3,6] are some of the phenomena discovered in this field. They all lead to spin-to-charge current interconversions, which are essential for future spin-orbit-based technological applications such as spin-orbit torques for magnetization switching [7, 8, 9, 10] or spin-orbit logic [11,12].

The SHE generates a transverse pure spin current when a charge current is applied in a material with strong SOC [1, 2]. Reciprocally, the inverse spin Hall effect (ISHE) is employed as a pure spin current detector because a measurable transverse charge current is created from the pure spin current. The spin-to-charge current conversion efficiency is given by the spin Hall angle, $\theta_{SH}$, which is the ratio between the spin Hall conductivity ($\sigma_{SH}$) and the longitudinal conductivity ($\sigma_{xx}$). There are several mechanisms that contribute to the SHE. The intrinsic contribution is described by the Berry curvature [13] and the extrinsic contributions, including skew scattering [14] and side jump [15], are caused by the impurities present in the host material. The strength of each mechanism reveals the potential of the material as a spin-to-charge converter and indicates the path to enhance $\theta_{SH}$. Determining the origin of the SHE in each material is thus mandatory for the development of efficient spin-orbit-based applications.

Heavy metals, such as Pt, Ta and W, are characterized by strongly spin-orbit-coupled bands and theory predicts that the intrinsic contribution should dominate the SHE [2, 16]. For the case of Pt, we have recently reported the crossover from the moderately dirty regime, where the intrinsic mechanism dominates, to the superclean regime, where the skew scattering governs [17]. The intrinsic $\sigma_{SH}$ value experimentally obtained in this work [1600±150 (ħ/e) Ω$^{-1}$cm$^{-1}$] is in good agreement with theoretical values [16, 18]. However, this is not the case for Ta.

Theoretical results suggest that β-Ta is characterized by a larger intrinsic $\sigma_{SH}$ [-378 (ℏ/e) $\Omega^{-1}cm^{-1}$] [19] than α-Ta [from -80 to -240 (ℏ/e) $\Omega^{-1}cm^{-1}$] [16, 19] and most experimental works are focused on β-Ta [20]. A large $\theta_{SH}$ (-7.5%, using $\sigma_{SH}$ in ℏ/e units [23]) was reported by Liu *et al.* for β-Ta [10], but discrepancies between values among different groups and techniques are common, ranging from -0.37% to -7.5% (using $\sigma_{SH}$ in ℏ/e units) in the literature [10, 19, 24, 30-33]. More importantly, there is no robust experimental evidence of which mechanism dominates the SHE in Ta, which hides the path to enhance $\theta_{SH}$.

In this Letter, we employ the spin absorption technique in lateral spin valves (LSVs) to study the SHE in Ta. We analyze a wide range of resistivities of Ta and observe the dominance of the intrinsic mechanism in the SHE of Ta, obtaining an intrinsic spin Hall conductivity, $\sigma_{SH}^{int}$, of -821±115 (ℏ/e) $\Omega^{-1}cm^{-1}$. We obtain the largest reported $\theta_{SH}$ for a pure metal, -35±3% (using $\sigma_{SH}$ in ℏ/e units [23]). The spin absorption method allows us to extract the spin diffusion length ($\lambda$) and $\theta_{SH}$ of Ta in the same device by making two independent magnetotransport measurements [17,24-28,34-36], in contrast to many other techniques, such as spin pumping [37], spin-torque ferromagnetic resonance [38], spin Hall magnetoresistance [39] or magneto-optical detection [29], where a thickness dependent study is carried out to assign an effective $\lambda$ and $\theta_{SH}$ to all samples, and not to each specific one. In addition, in the spin absorption technique, the spin Hall material is not in direct contact with the ferromagnet (FM) used as spin injector, which avoids proximity effects and magnetization dependent scattering [40]. Furthermore, temperature dependent measurements are easily accessible. With all these factors, we gain reliable insight into the mechanisms governing the SHE in Ta.

Different samples with Py/Cu lateral spin valves were fabricated on top of a $SiO_2$ (150nm)/Si substrate by using multiple-step e-beam lithography, subsequent metal deposition and lift-off. One of the LSVs contains a middle Ta wire and the other one, the reference LSV, does not. First, Py electrodes were patterned, with an interelectrode distance of 1 μm, and 30 nm of Py were e-beam evaporated at 0.6 Å/s and 1.4×10$^{-8}$ Torr. In the second step, 10 or 15 nm of Ta were sputtered at 1.6 Å/s, 20 W of power, 8×10$^{-8}$ Torr of base pressure and 1.5×10$^{-3}$ Torr of Ar pressure. In the third step, a 100-nm-wide channel was patterned and ~100 nm of Cu were thermally evaporated at 3 Å/s and 1.2×10$^{-8}$ Torr. In order to remove the ~2.4-nm-thick native oxide from the Ta wire [see Fig. 1(a)] and achieve highly transparent Ta/Cu and Py/Cu interfaces, the surfaces of the Py and Ta were *in-situ* cleaned with Ar-ion beam etching before the Cu deposition. Transmission electron microscopy (TEM) study was performed on a Titan 60-300 electron microscope (FEI Co., The Netherlands) equipped with imaging Cs corrector. High resolution TEM (HR-TEM) images were obtained at 300 kV at negative Cs imaging conditions [41]. The samples for TEM were fabricated by the standard focused ion beam (FIB) protocol [42]. All transport measurements described below were carried out in a $^4$He flow cryostat (applying external magnetic field $H$ and varying temperature $T$) using the lock-in technique (173 Hz and 575μA).

Structural characterization was performed in 10- and 15-nm-thick Ta films grown with the same conditions as the middle Ta wire. Figure 1(a) shows the cross section of the 15-nm-thick Ta film. The film is polycrystalline with seemingly random distribution of crystal orientations. As electron diffraction from such thin and laterally extended structure is technically difficult to obtain, we use its mathematical analog instead: 2D Fourier transform (FFT) of a high resolution image (Fig. 1(b)). The FFT pattern reveals coexistence of two types of reflections: with d~0.26 nm, which can be attributed to {002} lattice planes of β-Ta (either triclinic or hexagonal), and d~0.23 nm, which can be attributed to {110} lattice planes of cubic α-Ta [43]. It is remarkable that though α-Ta nanocrystals have random orientations, i.e. its reflections are uniformly distributed on the ring of the FFT pattern, the β-Ta phase shows a clear texture with the c-axis normal to the surface. Figure 1(c) combines virtual dark field images, reconstructed from the reflections marked on Fig. 1(b). The lateral size of α-Ta nanocrystals determined from this image is about 10 nm. At the same time, the β-Ta phase forms a continuous, yet heavily distorted, layer on top of α-Ta nanocrystals. Grazing incidence X-ray diffraction confirms the coexistence of β-Ta and α-Ta phases in both films [43]. The resistivity of these thin films is 209 μΩcm (10 nm) and 193 μΩcm (15 nm) at room temperature. Moreover, it shows a negative temperature coefficient of resistance (TCR) [43], which is characteristic of high-resistivity disordered metals and has been reported for the β-Ta phase [20, 44] and the amorphous Ta as well [21, 22].

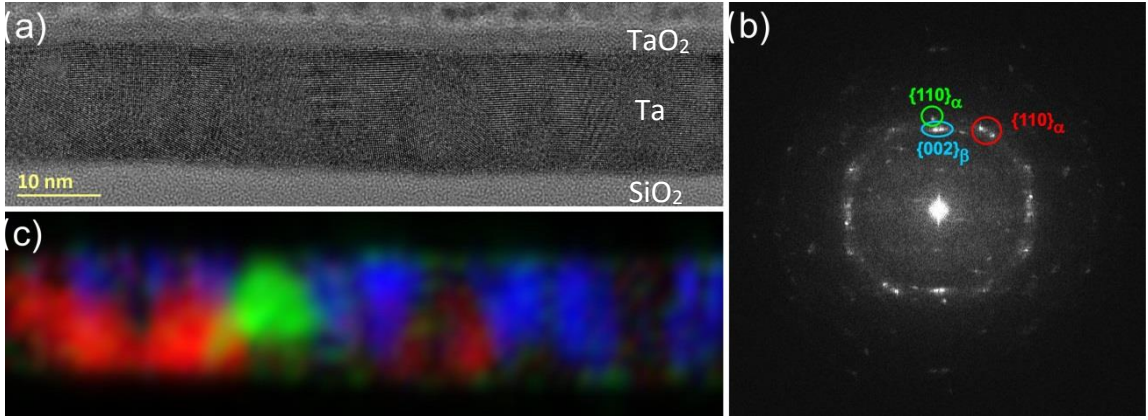

FIG. 1: (a) High resolution TEM image of a cross section of the 15-nm-thick Ta film. (b) FFT pattern of the layer on (a). (c) Superposition of color coded virtual dark field images reconstructed from reflections marked on (b) showing distribution, size and shape of β-Ta and α-Ta nanocrystals in the layer.

Figure 2(a) shows the SEM image of a LSV with a middle Ta wire. We inject a charge current from the Py electrode to the left side of the Cu channel. Due to the spin accumulation generated at the Py/Cu interface, spins diffuse to both sides of the Cu channel. On the right side, where no charge current is present, a pure spin current flows along the Cu channel, which is characterized by a long spin diffusion length [45-48]. The pure spin current will reach the second Py electrode, where the corresponding spin voltage is measured. This spin voltage is normalized to the injected current, obtaining the nonlocal resistance, $R_{NL}$, which depends on the relative magnetization of the two Py electrodes. The difference of the measured $R_{NL}$ at parallel and antiparallel configurations of the Py electrodes defines the spin signal, $\Delta R_{NL}$. The spin signal of the reference LSV, $\Delta R_{NL}^{ref}$, is larger than the one of the LSV with the middle Ta wire, $\Delta R_{NL}^{abs}$ (see Fig. 2(b) for the case of device D1). In the latter case, the Ta wire absorbs part of the spin current that flows along the Cu channel, so that the pure spin current reaching the detector is reduced. The ratio between the two spin signals is described by the one-dimensional spin diffusion model for transparent interfaces [17]:

$$\frac{\Delta R_{NL}^{abs}}{\Delta R_{NL}^{ref}} = \frac{2Q_{Ta}\left[\sinh\left(\frac{L}{\lambda_{Cu}}\right) + 2Q_{Py} e^{\left(\frac{L}{\lambda_{Cu}}\right)} + 2Q_{Py}^2 e^{\left(\frac{L}{\lambda_{Cu}}\right)}\right]}{\cosh\left(\frac{L}{\lambda_{Cu}}\right) - \cosh\left[\frac{L-2l}{\lambda_{Cu}}\right] + 2Q_{Py}\sinh\left[\frac{l}{\lambda_{Cu}}\right] e^{\frac{L-l}{\lambda_{Cu}}} + 2Q_{Ta}\sinh\left[\frac{L}{\lambda_{Cu}}\right] + 4Q_{Py}Q_{Ta}e^{\frac{L}{\lambda_{Cu}}} + 2Q_{Py}e^{\frac{l}{\lambda_{Cu}}}\sinh\left[\frac{(L-l)}{\lambda_{Cu}}\right] + 2Q_{Py}^2 e^{\frac{L}{\lambda_{Cu}}} + 4Q_{Py}^2 Q_{Ta}e^{\frac{L}{\lambda_{Cu}}}} \quad (1)$$

where $Q_{Py(Ta)} = \frac{R_{Py(Ta)}}{R_{Cu}}$, being $R_{Cu} = \frac{\lambda_{Cu}\rho_{Cu}}{w_{Cu}t_{Cu}}$, $R_{Py} = \frac{\lambda_{Py}\rho_{Py}}{w_{Cu}w_{Py}(1-\alpha_{Py}^2)}$ and $R_{Ta} = \frac{\lambda_{Ta}\rho_{Ta}}{w_{Cu}w_{Ta}tanh\frac{t_{Ta}}{\lambda_{Ta}}}$ the spin resistances of the Cu channel, Py electrodes and Ta wire, respectively. $\rho_{Cu,Py,Ta}$, $\lambda_{Cu,Py,Ta}$, $w_{Cu,Py,Ta}$ and $t_{Cu,Ta}$ are the resistivities, spin diffusion lengths, widths and thicknesses, respectively. The temperature dependence of $\rho_{Ta}$ for device D1 (see Fig. 2(c)) shows a negative TCR, as in the thin films [43]. $\alpha_{Py}$ is the spin polarization of Py. $L$ is the distance between the two Py electrodes, whereas $l$ is the distance between the Py injector and Ta wire. Since the spin resistances of Py and Cu are well known from our previous works [47, 48, 49], we can extract $\lambda_{Ta}$ from Eq. (1) [50]. We obtain $\lambda_{Ta}$=2.39±0.03 nm for device D1 at 10 K, which is in reasonable agreement with previous reports [24, 30-32, 52]. We repeated this procedure for different temperatures between 10 and 300 K.

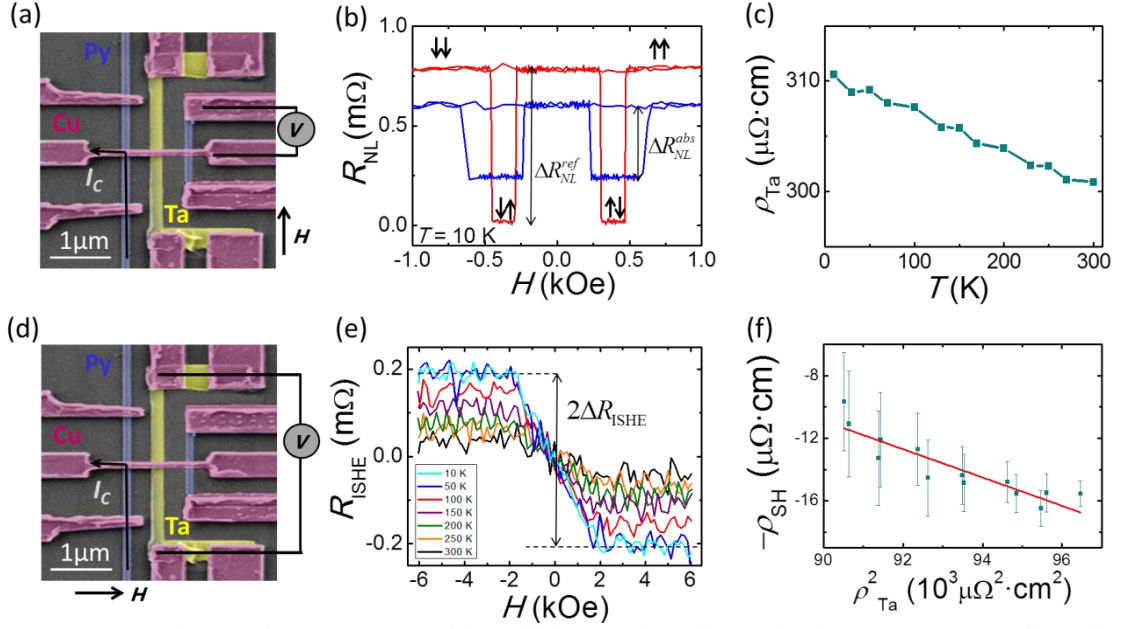

FIG. 2: (a) SEM image of a Py/Cu LSV with a middle Ta wire. The nonlocal measurement configuration and the orientation of the magnetic field are shown. (b) Nonlocal resistance as a function of magnetic field at 10 K measured in device D1 using the configuration shown in (a) for a Py/Cu LSV with (blue line) and without (red line) a Ta wire in between the Py electrodes. The reference spin signal ($\Delta R_{NL}^{ref}$) and the spin signal with Ta absorption ($\Delta R_{NL}^{abs}$) are tagged. (c) Temperature dependence of the resistivity of Ta in device D1. (d) SEM image of a Py/Cu LSV with a middle Ta wire. The ISHE measurement configuration and the orientation of the magnetic field are shown. (e) ISHE resistance as a function of the magnetic field at selected temperatures measured in device D1 using the configuration shown in (d). The ISHE signal ($2\Delta R_{ISHE}$) for 10 K is tagged. (f) Spin Hall resistivity as a function of the square of the longitudinal resistivity of Ta. Red solid line is the fitting of the data to Eq. (4).

Once demonstrated that Ta absorbs part of the spin current flowing along Cu, we measure the inverse spin Hall effect using the configuration shown in Fig. 2(d). By applying the magnetic field in plane, perpendicular to the easy axis of the Py electrodes, we inject a pure spin current into the Cu wire. This pure spin current will flow through the Cu channel and will partially be absorbed by the Ta wire, where it will be converted into a charge current due to the ISHE. The charge current generated along the Ta wire will be detected as a voltage drop. Normalizing the measured voltage drop with the injected current, we define the ISHE resistance, $R_{ISHE}$. If we reverse the magnetic field, the spins injected to the Cu channel will point in the opposite direction and, therefore, they will be deflected to the other side in the Ta wire, giving rise to the opposite $R_{ISHE}$, see Fig. 2(e) for the case of device D1. The difference between these two $R_{ISHE}$ values is twice the ISHE signal: $2\Delta R_{ISHE}$. Note that the obtained $\Delta R_{ISHE}$ in Ta is negative, i.e. opposite to the one obtained in Pt [17, 28] and Au [25, 28], as expected [16]. We measured $R_{ISHE}$ at different temperatures and, for the sake of clarity, only a selection is shown in Fig. 2(e). We observe that $\Delta R_{ISHE}$ decreases with increasing temperature.

The spin Hall resistivity, $\rho_{SH}$, a convenient magnitude to quantify the SHE, is related to $\theta_{SH}$ as $\theta_{SH} = -\rho_{SH}/\rho_{xx} = \sigma_{SH}/\sigma_{xx}$ [23], where $\rho_{xx}$ is the longitudinal resistivity ($\rho_{Ta}$ for Ta). The relation between $\Delta R_{ISHE}$ and $\rho_{SH}$ is given by [24]:

$$\rho_{SH} = -\frac{w_{Ta}}{x_{Cu,Ta}}\left(\frac{I_c}{\bar{I}_s}\right)\Delta R_{ISHE} \quad (2)$$

where $x_{Cu,Ta}$ is the shunting factor which takes into account the current in the Ta that is shunted through the Cu and is obtained from numerical calculations using a finite elements method [53]. $\bar{I}_s$ is the effective spin current that contributes to the ISHE in Ta and is given by [17, 28]:

$$\frac{I_s}{I_c} = \frac{\lambda_{Ta}\left(1-e^{-\frac{t_{Ta}}{\lambda_{Ta}}}\right)^2}{t_{Ta}\left(1-e^{-\frac{2t_{Ta}}{\lambda_{Ta}}}\right)} \cdot \frac{2\alpha_{Py}[Q_{Py}\sinh[(L-l)/\lambda_{Cu}]+Q_{Py}^2\, e^{(L-l)/\lambda_{Cu}}]}{\cosh\left(\frac{L}{\lambda_{Cu}}\right)-\cosh\left[\frac{L-2l}{\lambda_{Cu}}\right]+2Q_{Py}\sinh\left[\frac{l}{\lambda_{Cu}}\right]e^{\frac{L-l}{\lambda_{Cu}}}+2Q_{Ta}\sinh\left[\frac{L}{\lambda_{Cu}}\right]+4Q_{Py}Q_{Ta}e^{\frac{l}{\lambda_{Cu}}}+2Q_{Py}e^{\frac{l}{\lambda_{Cu}}}\sinh\left[\frac{(L-l)}{\lambda_{Cu}}\right]+2Q_{Py}^2 e^{\frac{L}{\lambda_{Cu}}}+4Q_{Py}^2 Q_{Ta} e^{\frac{L}{\lambda_{Cu}}}}. \quad (3)$$

Note that quantifying $\rho_{SH}$ requires the value of $\lambda_{Ta}$ that we extract from the spin absorption experiment.

Phenomenologically, each mechanism of the SHE contributes with a different resistivity dependence to the $\rho_{SH}$ [54]:

$$-\rho_{SH} = \sigma_{SH}^{int}\rho_{Ta}^2 + \sigma_{SH}^{sj}\rho_{Ta,0}^2 + \alpha_{ss}\rho_{Ta,0} \quad (4)$$

where $\sigma_{SH}^{int}$ is the intrinsic spin Hall conductivity of Ta, $\sigma_{SH}^{sj}$ is the one corresponding to the side jump, $\alpha_{ss}$ is the skew scattering angle and $\rho_{Ta,0}$ is the residual resistivity of Ta. In order to unveil the weight of each mechanism, we analyze the dependence of -$\rho_{SH}$, obtained experimentally using Eq. (2) and (3), with the longitudinal resistivity of Ta. In Fig. 2(f), we plot -$\rho_{SH}$ against $\rho_{Ta}^2$ and fit this data to a linear function with the slope that corresponds to $\sigma_{SH}^{int}$ and the intercept to the sum of the skew scattering and side jump contributions. We extract $\sigma_{SH}^{int}$=–910±130 (ℏ/e) Ω⁻¹cm⁻¹ and $\sigma_{SH}^{sj}\rho_{Ta,0}^2 + \alpha_{ss}\rho_{Ta,0}$=71±12 (e/ℏ) μΩ·cm from device D1.

The variation of $\rho_{Ta}$ with temperature is very small, around 3% as shown in Fig. 2(c), thus the studied resistivity range is relatively short. In order to get a more complete study, covering a broader range of resistivities, we use additional devices (D2-D7) containing Ta wires with different $\rho_{Ta,0}$ (see Table I). In order to tune $\rho_{Ta,0}$, we modify the width and thickness of the Ta wires. For each device, we measure first the spin absorption signal at 10 K, in order to extract $\lambda_{Ta}$ of each Ta wire (see Table I). We observe that $\lambda_{Ta}$ is small for all devices, between 0.8 and 2.4 nm, and has no clear tendency with the resistivity.

Next, we measure the ISHE for each device at 10 K. Figure 3(a) shows $R_{ISHE}$ at 10 K for three selected devices that are characterized by different $\rho_{Ta,0}$. We observe that |Δ$R_{ISHE}$| increases with $\rho_{Ta,0}$. This result is consistent for all the studied devices, as shown in the inset of Fig. 3(a). Using Eqs. (2) and (3), we extract the $\rho_{SH}$ for each device, which is plotted in Fig. 3(b) as a function of $\rho_{Ta,0}$. A clear increase of |$\rho_{SH}$| with $\rho_{Ta,0}$ is observed.

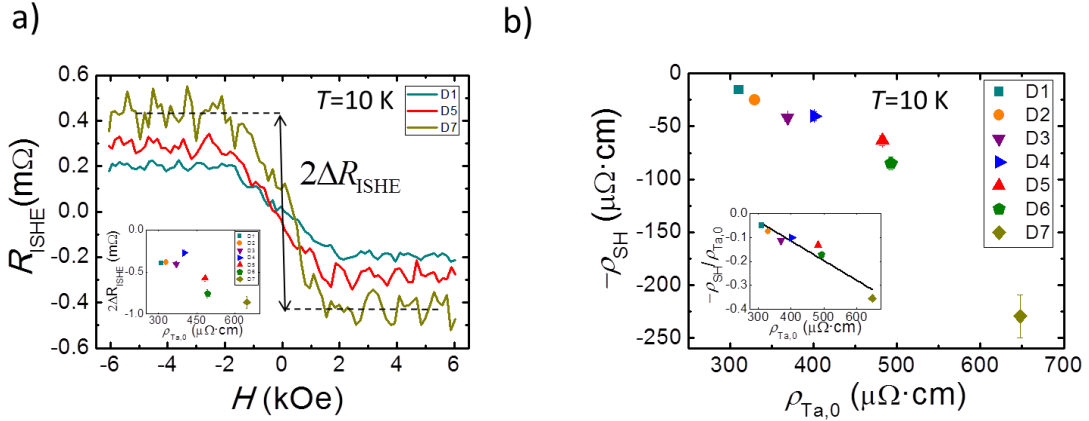

FIG. 3: (a) ISHE resistance as a function of the magnetic field at 10 K for selected devices measured using the configuration shown in Fig. 2(d). Inset: ISHE signal as a function of the residual resistivity of Ta for all devices at 10 K. (b) Spin Hall resistivity as a function of the residual resistivity of Ta for all devices at 10 K. Inset: Ratio of the spin Hall resistivity and residual resistivity of Ta as a function of the residual resistivity of Ta for all devices at 10 K. Black solid line is the fitting of the data to Eq. (4) at low temperature.

Since, at low temperature, Eq. (4) can be rewritten as $\frac{-\rho_{SH}}{\rho_{Ta,0}} = (\sigma_{SH}^{int} + \sigma_{SH}^{sj})\rho_{Ta,0} + \alpha_{ss}$, we can perform a linear fit of $\frac{-\rho_{SH}}{\rho_{Ta,0}}$ against $\rho_{Ta,0}$. Using the experimental data of all devices, see the inset of Fig. 3(b), we extract $\sigma_{SH}^{int} + \sigma_{SH}^{sj}$=−820±120 (ℏ/e) $\Omega^{-1}cm^{-1}$ from the slope and $\alpha_{ss}$=0.21±0.05 from the intercept. The obtained $\sigma_{SH}^{int}$ from device D1 as a function of temperature ($\sigma_{SH}^{int}$=−910±130 (ℏ/e) $\Omega^{-1}cm^{-1}$), is compatible with the $\sigma_{SH}^{int} + \sigma_{SH}^{sj}$= −820±120 $\Omega^{-1}cm^{-1}$ result obtained considering all devices. This indicates that $\sigma_{SH}^{sj}$ is negligible in our devices, which is expected in a pure metal [1, 55, 56]. Therefore, from the previous result in device D1 as a function of temperature, we can consider $\alpha_{ss}\rho_{Ta,0}$=71±12 (e/ℏ) μΩ·cm, which leads to $\alpha_{ss}$=0.23±0.04. This skew scattering angle is also consistent with the last result, $\alpha_{ss}$=0.21±0.05, obtained using all devices at low temperature.

Considering that the upper part of the Ta wire, where the spin absorption from Cu is occurring, is composed by β-Ta grains, and that the spin diffusion length of Ta is of few nanometers, we can safely consider that the spin-to-charge conversion occurs in the upper β-Ta grains. Therefore, the obtained $\sigma_{SH}^{int}$, -820±120 (ℏ/e) $\Omega^{-1}cm^{-1}$, is dominated by β-Ta. In the literature we cannot identify reliable quantitative predictions for a polycrystalline sample including β-Ta. In Ref. 19 the authors discuss the transition in a binary $Ta_{1-x}Au_x$ alloy using the rigid band (virtual crystal) approximation to model doped Ta at some level. Furthermore, the authors present numerical results for β-Ta (-240 (ℏ/e) $\Omega^{-1}cm^{-1}$) but in contrast to α-Ta they do not present results changing the Fermi energy. However, in a system such as β-Ta it can be expected that $\sigma_{SH}^{int}$ changes dramatically as a function of the Fermi energy as shown in Ref. 57 for β-W. Using Fig. 2 of Ref. 57 and assuming the virtual crystal approximation, going from β-W to β-Ta would reduce the Fermi energy by ~1.3eV and lead $\sigma_{SH}^{int}$ quantitatively close to the value identified in this work. It further highlights that in this region $\sigma_{SH}^{int}$ changes dramatically, inducing large changes for small variations in the Fermi energy. It is remarkable that not only the intrinsic contribution is significant in our Ta wires, but also the extrinsic one. The obtained $\alpha_{ss}$=0.21±0.05 corresponds to an extrinsic contribution of $\theta_{SH}$=21% independent of the residual resistivity. Nevertheless, due to the opposite signs of the contributions and the high resistivity, the skew scattering is counterbalanced by the intrinsic contribution, which becomes dominant.

Table I: Thickness ($t_{Ta}$), width ($w_{Ta}$), resistivity ($\rho_{Ta}$), spin diffusion length ($\lambda_{Ta}$) and spin Hall angle ($\theta_{SH}$) of the Ta wires obtained in this work using the spin absorption technique in lateral spin valves. The shunting factor ($x_{Cu,Ta}$) of the different devices is shown. All data correspond to 10 K.

| Device | $t_{Ta}$(nm) | $w_{Ta}$(nm) | $\rho_{Ta}$(μΩ·cm) | $\lambda_{Ta}$(nm) | $\theta_{SH}$ (%) | $x_{Cu,Ta}$ |
|---|---|---|---|---|---|---|
| D1 | 15 | 270 | 311 | 2.39±0.03 | -5.0±0.3 | 0.09547 |
| D2 | 15 | 270 | 330 | 1.27±0.02 | -7.6±0.6 | 0.09556 |
| D3 | 15 | 270 | 369 | 0.81±0.02 | -11.3±0.9 | 0.10198 |
| D4 | 10 | 273 | 401 | 1.52±0.05 | -10±1 | 0.04106 |
| D5 | 10 | 224 | 483 | 1.31±0.02 | -13.2±0.9 | 0.04451 |
| D6 | 10 | 187 | 493 | 2.22±0.06 | -17±1 | 0.04096 |
| D7 | 10 | 195 | 648 | 0.76±0.03 | -35±3 | 0.03868 |

Figure 4 shows the absolute value of the spin Hall angle of Ta as a function of the resistivity of Ta, together with the one of Pt obtained in our previous work [17]. We are able to increase linearly $\theta_{SH}$ of Ta up to -35±3% by simply increasing the resistivity of the Ta wire (following the definition used by other groups, in units of ℏ/2e, our $\theta_{SH}$ is -70% [23]). This is a clear indication of the dominance of the intrinsic mechanism in the SHE of Ta. We observed a similar tendency in Pt in the intrinsic regime, but with a larger slope, as shown in Fig. 4, due to the larger $\sigma_{SH}^{int}$ in Pt than in Ta.

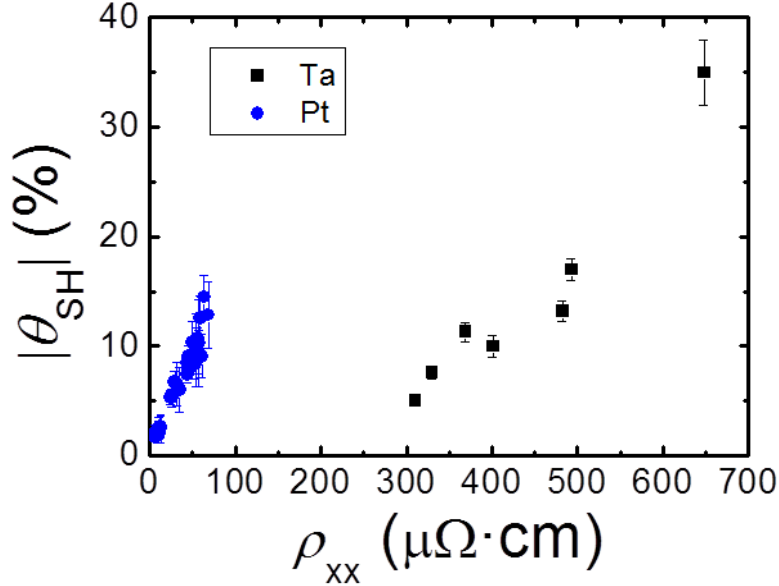

FIG. 4: Longitudinal resistivity dependence of the absolute value of the spin Hall angle of Ta (black solid squares) and Pt (blue solid circles, from Ref. 17).

To conclude, we experimentally determine the intrinsic mechanism as the leading contribution of the SHE in highly-resistive Ta. We extract $\sigma_{SH}^{int}$ for β-Ta to be -820±120 (ℏ/e) $\Omega^{-1}cm^{-1}$, which is constant in a broad range of resistivities. The predominance of the intrinsic mechanism reveals the path to increase the spin Hall angle in Ta: increasing the resistivity of the metal. With this approach, by measuring with the spin absorption technique, we can systematically vary $\theta_{SH}$ from -5.0±0.3% up to -35±3%, achieving the largest conversion efficiency reported so far for a pure metal. This work unveils the intrinsic potential of Ta as a spin-to-charge converter, definitely appealing and promising for spin-orbit-based technological applications.

**Acknowledgements**

This work is supported by the Spanish MINECO under the Maria de Maeztu Units of Excellence Programme (MDM-2016-0618) and under Projects No. MAT2015-65159-R and MAT2017-82071-ERC, by Semiconductor Research Corporation under Project No. 2017-IN-2744 and by the Japanese Grant-in-Aid for Scientific Research on Innovative Area, "Nano Spin Conversion Science" (Grant No. 26103002). M.G. acknowledges financial support from the Leverhulme Trust via an Early Career Research Fellowship (ECF-2013-538). E.S thanks the Spanish MECD for a Ph.D. fellowship (Grant No. FPU14/03102). Y. Omori acknowledges financial support from JSPS through "Research program for Young Scientists" and "Program for Leading Graduate Schools (MERIT)".

# SUPPLEMENTAL MATERIAL

## S1. Additional structural characterization of Ta thin films

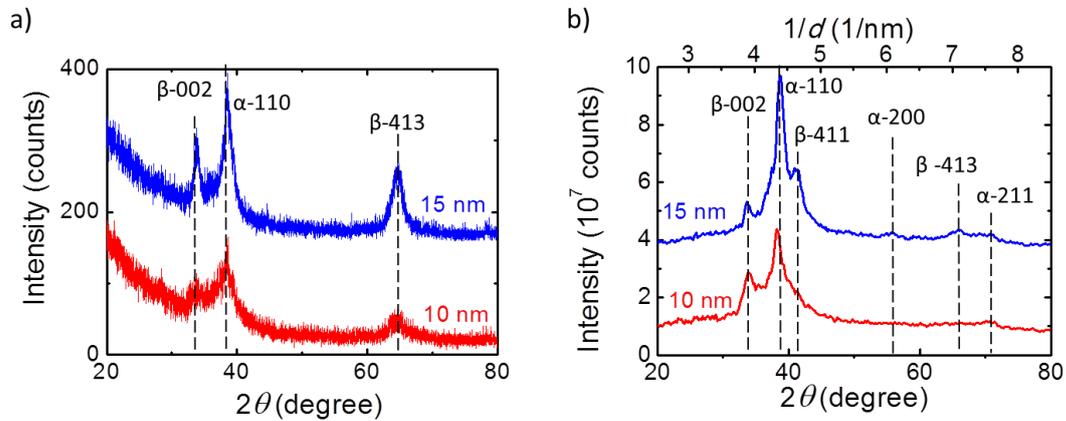

FIG. S1: a) Grazing incidence X-ray diffraction 2Θ scan for a grazing incidence angle of $\phi=0.5°$ in 15- (blue line) and 10-nm-thick (red line) Ta thin films. The data of 15-nm-thick Ta has been shifted for clarity. b) Electron diffraction pattern obtained by TEM for 15- (blue line) and 10-nm-thick (red line) Ta thin films. The data of 15-nm-thick Ta has been shifted for clarity.

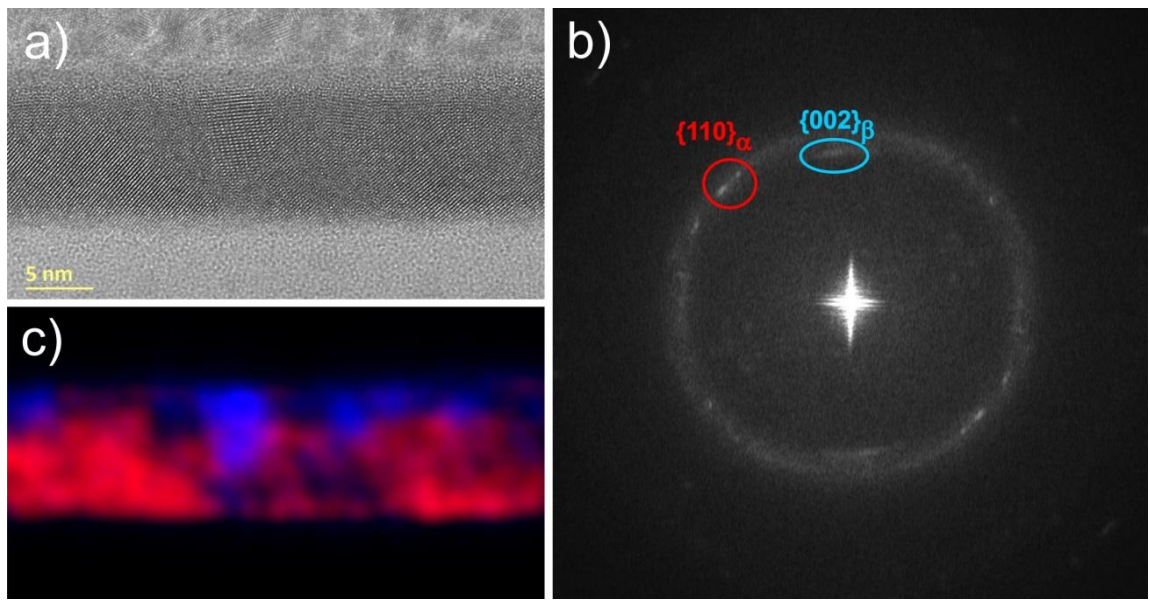

FIG. S2: (a) High resolution TEM image of a cross section of the 10-nm-thick Ta film. (b) FFT pattern of the layer on (a). (c) Superposition of color coded virtual dark field images reconstructed from reflections marked on (b) showing distribution, size and shape of β-Ta and α-Ta nanocrystals in the layer.

## S2. Temperature dependence of the resistivity in Ta thin film and Ta nanowire.

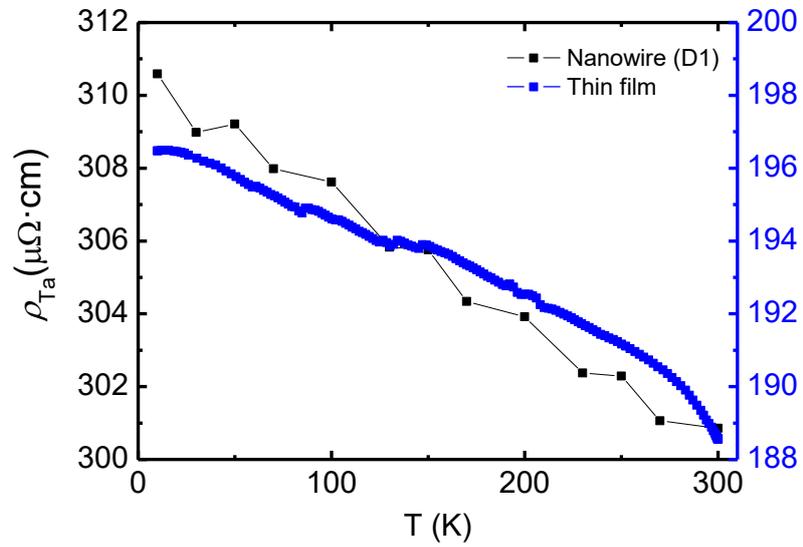

FIG. S3: Temperature dependence of the resistivity in the 15-nm-thick Ta thin film (blue line) and 15-nm-thick Ta nanowire of Device D1 (black line). We observe that in both cases the variation of $\rho_{Ta}$ with temperature is similar, around ~ 3-4%.